\documentclass{article}

\begin{document}

\title{Wave-like spatially homogeneous models\\
of Stackel spacetimes (3.1) type\\
in the scalar-tensor theory of gravity}

\author{Evgeny~Osetrin\thanks{evgeny.osetrin@gmail.com}, \\[1ex]
Konstantin~Osetrin\thanks{osetrin@tspu.edu.ru},  \\[1ex]
Altair~Filippov\thanks{altair@tspu.edu.ru},  \\[1ex]
Ilya~Kirnos\thanks{ikirnos@sibmail.com} \\[2ex]
Center for Theoretical Physics, Tomsk State Pedagogical University, \\[1ex]
Kievskaya str., 60, Tomsk, 634061, Russia\\
}

\maketitle

\begin{abstract}
All classes of spatially homogeneous space-time models in the generalized scalar-tensor theory of gravity are found that allow the integration of the equations of motion of test particles and the eikonal equation by the method of %complete
separation of variables by type (3.1). Three classes of exact solutions are obtained that relate to wave-like spacetime models. The resulting spacetime models are of type IV, VI, and VII according to the Bianchi classification and type N according to Petrov's classification.
\end{abstract}

%\keywords{scalar-tensor theory of gravity; wave-like spacetimes; spatially homogeneous spacetimes; exact solutions.}

\section{Introduction}	

Spacetime models that allow the existence of privileged coordinate systems with respect to which the equation of motion of test particles in the form of the Hamilton-Jacobi equation allows integration by the method of complete separation of variables is called Stackel  spaces 
(in honor of Paul St\"{a}ckel  \cite{Stackel}). Details of the theory of the Stackel spaces can be found 
%in \cite{Shapovalov1}-\cite{Obukhov}. 
in \cite{Shapovalov1,Shapovalov2,Obukhov}. 
Among the Stackel spaces in the four-dimensional case, there are three types of spaces that allow separation of isotropic (null/wave) variables. We will call such spaces wave-like.
These space-time models make it possible to integrate the equations of motion of test particles and radiation and obtain exact solutions of field equations in metric theories of gravity (see \cite{Bagrov,OFO,OO} for examples).
The simplest type of wave-like Stackel spaces are spaces of type (3.1) that admit three commuting Killing vector fields. In a privileged coordinate system, the metric of spaces of type (3.1) depends on only one isotropic (wave) variable and can be written in the form:
\begin{equation}
dS^2=g_{ij}(x^0)\,dx^idx^j,
\qquad
% g_{ij}=g_{ij}(x^0),
%\qquad
 g_{00}=0,
\qquad
i,j,k = 0...3,
\label{metric0}
\end{equation}
where $x^0$ is a null (wave) variable, i.e. 
$dS=0$ for $dx^0\ne 0$ and $dx^r=0$ ($ r=1...3$).

The metric (\ref{metric0}) can be represented in the following general form:
\begin{equation}
dS^2=2dx^0dx^1+g_{ab}(x^0)\Big(dx^a+{f}^a(x^0)dx^1\Big)\Big(dx^b+{f}^b(x^0)dx^1\Big),
\quad
a,b =2,3.
\label{metric1}
\end{equation}

In this paper, we consider the scalar-tensor theory of gravitation of a general form with the Lagrangian (see \cite{Odintsov1,Odintsov2,Capozziello})
\begin{equation} 
L=\frac{1}{16\pi}\sqrt{-g}\left[  \phi R -\frac{\omega (\phi)}{\phi}\nabla^k\phi\nabla_k\phi
-2\Lambda(\phi)  \right] + L_{m}.
\label{Lagr}
\end{equation}
The field equations of the scalar-tensor theory of gravity (\ref{Lagr}) in vacuum take the form:
\begin{equation} 
\phi G_{ij}+\left( \nabla^k\nabla_k \phi+\frac 12\frac{\omega (\phi)}{\phi}\nabla^k\phi\nabla_k\phi+\Lambda  (\phi)\right)g_{ij}-\nabla_i\nabla_j\phi-\frac\omega\phi\nabla_i\phi\nabla_j\phi=0,
%\epsilon l_il_j,
\label{eq2}
\end{equation}
\begin{equation} 
(2\omega+3)\nabla^k\nabla_k \phi+\omega'(\phi)\nabla^k\phi\nabla_k\phi+4\Lambda(\phi)-2\phi \Lambda'(\phi)=0.
\label{eq3}
\end{equation}
where ${\omega (\phi)}$ and $\Lambda(\phi)$  are, in the general case, arbitrary functions of the scalar field. % $\phi$.
 
When considering vacuum models of spacetime with a cosmological constant both in Einstein’s theory of gravity and in the generalized scalar-tensor theory of gravitation for the metric (\ref{metric1}), it can be shown that the functions $ {f}^a (x^0) $ and the cosmological constant (in the scalar tensor theory the function $ \Lambda (\phi) $ plays the role of the cosmological constant) vanish, and the metric takes on a simpler form:
\begin{equation}
dS^2=2\,dx^0 dx^1+g_{ab}(x^0)\,dx^a\,dx^b,
\qquad
a,b =2,3.
\label{metric2}
\end{equation}
For the metric (\ref{metric2}) there remains only one nonnull component of the Ricci tensor -  $R_{00}$, that does not vanish identically. The scalar curvature $R$ also vanishes.

Spatially homogeneous spacetime models play an important role in constructing models for the development of the Universe. 
From this point of view, in the study of spatially homogeneous models of interest are both the study of permissible types of gravitational waves, and the possibility of analytical integration in these models of the eikonal equation and the equation of motion of test particles.
Earlier, in \cite{Osetrin1, Osetrin2}, we studied the conditions under which the considered wave-like spacetime models are spatially homogeneous.
We have previously obtained \cite{Osetrin3, Osetrin4} for the metric (\ref{metric2}) three possible classes of spatially homogeneous spacetime models, corresponding to types IV, VI, and VII, respectively, according to the Bianchi classification and type N according to Petrov's classification.

In this paper, we obtain for the models under consideration exact solutions of the field equations of the generalized scalar tensor theory of gravity 
(\ref{eq2})-(\ref{eq3}). % and discuss the differences from similar models in Einstein's theory of gravity.

\section{Spatially homogeneous wave-like Stackel (3.1) spacetimes}

Note that for all spatially homogeneous models considered below with the metric (\ref{metric2}), the field equations (\ref{eq2})-(\ref{eq3}) 
lead to $ \Lambda = 0 $ and reduce to a single equation on scalar field of the form:
\begin{equation}
\phi ''({x^0}) +
\frac{\omega (\phi)}{\phi ({x^0})} \, \phi '({x^0})^2
%- \frac{{k}}{4}
+\frac{{k} \,\phi ({x^0})}{ 4\, {x^0}^2}
=0,
\quad
{k} - \mbox{const}.
\label{eq4}
\end{equation}
The value of the constant $ {k} $ is determined by the parameters of
specific model.

For the Brans-Dicke theory, when $\omega=\mbox{const}$, for the scalar field $\phi ({x^0})$  from Eq.~(\ref{eq4}) we get:
\begin{equation}
\phi ({x^0})=\left( c_1 {x^0}^{\left(1-{\mu}\right)/2}+c_2 {x^0}^{\left(1+{\mu}\right)/2}\right)^{1/({1+\omega})},
\quad
\omega, {\mu}, c_1, c_2 - \mbox{const},
\label{scalar}
\end{equation}
where
$
{\mu}=\sqrt{1-{k}\, (\omega +1)}
$,
and
\begin{equation}
1- {k}\, (\omega +1)\ge 0.
\end{equation}
The models of spacetime considered admits three commuting Killing vectors
 $X^i_{(p)}$:
 % ($p,q=1,2,3$):
 \begin{equation}
X_{(0)}^i= 
\left(0,\, 1,\,  0,\,  0\right),
\qquad
 X_{(1)}^i= 
\left(0,\, 0,\,  1,\,  0\right),
\qquad
 X_{(2)}^i= 
\left(0,\, 0,\,  0,\,  1\right).
\end{equation}
The vector $ X_{(0)}^i$ is a null vector, and the vectors $ X_{(1)}^i$ and  $X_{(2)}^i$ are spacelike vectors.

The additional Killing vector, which can provides the spatial homogeneity of the models, must have the following form:
$$
 X_{(3)}^i= 
\left(-x^0,\,  x^1,\, {p} x^2+{q} x^3,\,  \tilde{p} x^2+\tilde{q} x^3\right)
$$
where ${p}$, $\tilde{p}$, ${q}$, $\tilde{q}$
%, $\gamma$, $\delta$ 
are constants.

Then Killing vectors $X_{(1)}^i$, $X_{(2)}^i$, $X_{(3)}^i$ can provide spatial homogeneity of the model. 
Below we consider three classes of emerging models.

\subsection{Spatially homogeneous wave-like model {I} type (Bianchi~VI)}

The additional Killing vector for this class of spacetimes has the following form:
\begin{equation}
 X_{(3)}^i= 
\left(-x^0,\,  x^1,\,  {p} x^2,\,  {q} x^3\right),
\end{equation}
where $x^0$ is a null wave-like variable and ${p}$, ${q}$ -- constants.

%\vspace{2ex}

The commutation relations for the Killing vectors defining a subgroup of spatial isometry in the model {I} have the form:
\begin{equation} 
\big[  X_{(1)}, X_{(2)} \big]=  0, 
\qquad
\big[  X_{(1)}, X_{(3)} \big]=  {p} X_{(1)}, 
\qquad
\big[  X_{(2)}, X_{(3)} \big]= {q}  X_{(2)}.
%\label{Ieq3}}
\end{equation} 

The metric of model {I} can be written as
\begin{equation} 
{\rm g}_{ij}= \left(                                                                                                                            
              \begin{array}{cccc}
               0 & 1 & 0 & 0 \\
               1 & 0 & 0 & 0 \\[1ex]
               0 & 0 & {{x^0}^{2 {p}}}/{{\sigma}^2} & -{{\alpha} {x^0}^{({p}+{q})}}/{{\sigma}^2} \\[1ex]
               0 & 0 & -{{\alpha} {x^0}^{({p}+{q})}}/{{\sigma}^2} & {{x^0}^{2 {q}}}/{{\sigma}^2}
              \end{array}
              \right),
\end{equation} 
where $\alpha$, %$\sigma$, 
${p}$, ${q}$ are independent constant parameters of model {I}, $x^0$ is a null (wave) variable and
$$
%{g=\det g_{ij}=}-\frac{{x^0}^{2 ({p}+{q})}}{{\sigma}^2},
%\qquad
{\sigma}^2=1-{\alpha}^2,
\qquad
-1<{\alpha}<1,
\qquad
0<{\sigma}^2\le 1.
$$
The scalar field $ \phi $ is defined in the general case by Eq.~(\ref{eq4}) or, in the case of the Brans-Dicke theory, by the relation (\ref{scalar}), where constant $ {k} $ for the model~{I} has the form:
\begin{equation} 
{k}=2\,\frac{\left({\alpha}^2-2\right) {p}^2+2 {p} \left({\alpha}^2 ({q}-1)+1\right)+{q} \left({\alpha}^2 ({q}-2)-2 {q}+2\right)}{\left({\alpha}^2-1\right)}.
\end{equation} 
In the general case, there remain three independent non-zero components of the Weyl tensor ${\rm C}_{0202}$,  
${\rm C}_{0303}$ and ${\rm C}_{0203}$.
Note that for each value of $\alpha\ne 0$ there are 2 pairs of values of ${p}$, ${q}$ for which ${\rm C}_{0202}$ can be turned to zero, while ${\rm C}_{0303}$ will not be zero  and vice versa.
For example: for $\alpha=\pm 1/\sqrt 2$,  ${p} =1/2$ and ${q}=(3\pm 2\sqrt 3)/6$ we have ${\rm C}_{0303}=0$, but ${\rm C}_{0202}\ne 0$.
Spacetime of the model {I} becomes flat in the following  three cases only:
\begin{equation}
%A. \,{p}={q}\,  (\mbox{then }{p},{q}\mbox{ are }0\mbox{ or }1); 
\mbox{A.} \,{p}={q}\,\,  ({p},{q}\mbox{ are }0\mbox{ or }1); 
\quad
%2.\quad\mbox{}&& {p}={q}=1,\\
\mbox{B.} \,{\alpha}=0, \, {p}=0,\, {q}=1;
\quad
\mbox{C.} \,{\alpha}=0, \, {p}=1,\, {q}=0.
\end{equation}
Non-flat spacetimes of model {I} are of type VI${}_a$ according to Bianchi's classification %($a=|({p}+{q})/({p}-{q}|$)
 and type N according to Petrov's classification.

\subsection{Spatially homogeneous wave-like model {II} type (Bianchi~IV)}

The additional Killing vector for model {II} %of spacetimes 
has the following form:
\begin{equation}
 X_{(3)}^i= 
\left(-x^0,\,  x^1,\,  {p} x^2,\,  x^2+{p} x^3\right),
\end{equation}
where $x^0$ is a null wave-like variable, ${p}$ is a constant.

The commutation relations for the model {II} have the form:
\begin{equation} 
\big[  X_{(1)}, X_{(2)} \big]=  0, 
\qquad
\big[  X_{(1)}, X_{(3)} \big]= {p} X_{(1)}+ X_{(2)}, 
\qquad
\big[  X_{(2)}, X_{(3)} \big]= {p}  X_{(2)}.
\label{IIeq3}
\end{equation}

Metric of model {II} % in the preferred coordinate system
can be written as
\begin{equation}
{\rm g}_{ij}= \left(                                                                                                                                                                                                                                 
              \begin{array}{cccc}
               0 & 1 & 0 & 0 \\
               1 & 0 & 0 & 0 \\
               0 & 0 & {{x^0}^{2 {p}} \left({{\alpha^2}} \ln  ^2{x^0}-2 {{\beta}} \ln {x^0}+{{\gamma^2}}\right)}/{{\sigma}^2} & {{x^0}^{2 {p}} ({{\alpha^2}} \ln  {x^0}-{{\beta}})}/{{\sigma}^2} \\
               0 & 0 & {{x^0}^{2 {p}} ({{\alpha^2}} \ln  {x^0}-{{\beta}})}/{{\sigma}^2} & {{{\alpha^2}} {x^0}^{2 {p}}}/{{\sigma}^2}
              \end{array}
              \right)
\end{equation}
where $\alpha$, $\beta$, $\gamma$,
${p}$ are independent constant parameters of model, $x^0$ is a null wave-like variable and
\begin{equation}
%{g=\det g_{ij}=}-\frac{{x^0}^{4 {p}}}{{\sigma}^2},
%\qquad
{{\sigma}^2}={{\alpha^2}}{{\gamma^2}} - {{\beta}}^2,
\qquad
{\alpha}{\gamma}{\sigma}\ne 0,
\qquad
 0\le {{\beta}}^2<({{\alpha}}{{\gamma}})^2.
\end{equation}
The scalar field $ \phi $ is defined in the general case by Eq.~(\ref{eq4}) or, in the case of the Brans-Dicke theory, by the relation (\ref{scalar}), where constant $ {k} $ for the model~{II} has the form:
\begin{equation} 
{k}=2\,\frac{4 {\sigma}^2 ({p}-1) {p}+{\alpha}^2}{{\sigma}^2 }.
\end{equation} 
Rieman tensor and Weyl tensor of the model II can not be equal to zero.
%
%Classification of spacetimes by Bianchi --  type IV.
The spacetimes of the model {II} are of type IV %${}_a$ 
according to Bianchi's classification %($a=$) 
and type N according to Petrov's classification.

\subsection{Spatially homogeneous wave-like model {III} type (Bianchi~VII)}

The additional Killing vector for this model of spacetimes has the following form:
\begin{equation}
 X_{(3)}^i= 
\left(-x^0,\,  x^1,\,  {p} x^2-x^3,\,  x^2+{p} x^3\right),
\end{equation}
where $x^0$ is a null wave-like variable, ${p}$ is a constant.

The commutation relations for the model {III} have the form:
$$
\big[  X_{(1)}, X_{(2)} \big]=  0, 
\qquad
\big[  X_{(1)}, X_{(3)} \big]=  {p} X_{(1)}+ X_{(2)}, 
$$
\begin{equation} 
\big[  X_{(2)}, X_{(3)} \big]= -X_{(1)}+ {p}  X_{(2)}.
\label{IIIeq3}
\end{equation}

The metric of the model {III} %${\rm g}_{ij}$
have a form:
%\begin{equation} 
%{\rm g}_{ij}=
% \left(                                                                                                                                                                                                                                          
%              \begin{array}{cccc}
%               0 & 1 & 0 & 0 \\
%               1 & 0 & 0 & 0 \\[1ex]
%               0 & 0 &{{x^0}^{2 {p}}\big({\gamma}-{}{\alpha} \cos ( \ln{x^0}^2)-{}{\beta} \sin ( \ln{x^0}^2){} \big)}/{{\sigma}^2} & {{x^0}^{2 {p}}\big ({}{\alpha} \sin (\ln{x^0}^2)-{}{\beta} \cos ( \ln{x^0}^2)\big)}/{{\sigma}^2} \\[1ex]
%               0 & 0 & {{x^0}^{2 {p}} \big({}{\alpha} \sin ( \ln{x^0}^2)-{}{\beta} \cos ( \ln{x^0}^2)\big)}/{{\sigma}^2} & {{x^0}^{2 {p}} \big({\gamma}+{}{\alpha} \cos ( \ln{x^0}^2)+{}{\beta} \sin ( \ln{x^0}^2){}\big)}/{{\sigma}^2}
%              \end{array}
%              \right)
%\end{equation}
\begin{eqnarray}
dS^2 &=& 
2\,dx^0dx^1+
\frac{{x^0}^{2 {p}}}{{\sigma}^2}\,
\Bigl(
{\big({\gamma}-{}{\alpha} \cos ( \ln{x^0}^2)-{}{\beta} \sin ( \ln{x^0}^2){} \big)}\,{dx^2}^2 \nonumber \\
& & 
\mbox{}+
2\,{\big ({}{\alpha} \sin (\ln{x^0}^2)-{}{\beta} \cos ( \ln{x^0}^2)\big)}\,dx^2dx^3 \nonumber \\
& &
\mbox{}+
{ \big({\gamma}+{}{\alpha} \cos ( \ln{x^0}^2)+{}{\beta} \sin ( \ln{x^0}^2){}\big)}\,{dx^3}^2
\Bigr),
\end{eqnarray}
where $\alpha$, $\beta$, $\gamma$,
${p}$ are constant parameters of the model %, $x^0$ is a null wave-like variable 
and
$$
\sigma^2={\gamma}^2- {\alpha}^2-{\beta}^2,
\qquad
{\gamma}\ne 0,
\qquad
{\gamma}^2> {\alpha}^2+{\beta}^2.
$$
The scalar field $ \phi $ is defined in the general case by Eq.~(\ref{eq4}) or, in the case of the Brans-Dicke theory, by the relation (\ref{scalar}), where constant $ {k} $ for the model~{III} has the form:
\begin{equation} 
{k}=8\, \frac{{\alpha}^2+{\beta}^2+({p}-1) {p} {\sigma}^2}{{\sigma}^2}.
\end{equation} 
The model {III} becomes flat only if
$
{\alpha}={\beta}=0
$
(then
$
{p}=0 \mbox{ or } 1
$).
The spacetimes of model {III} are of type VII${}_a$  according to Bianchi's classification % ($a={p}$) 
and type N according to Petrov's classification.

\section{Conclusion}

A classification of spatially homogeneous plane-wave models of spacetime in tensor-scalar theory of gravity is constructed.
Three classes of wave-like spatially homogeneous exact models of spacetime  in tensor-scalar theory are obtained. 
% ({I}, {II}, {III}).
%The models considered can describe the primordial gravitational waves of the Universe.
The models considered could describe  the primordial periodic and aperiodic gravitational waves in a spatially homogeneous Universe.

\section*{Acknowledgments}

The reported study was funded by RFBR, project number N~20-01-00389~A.

%\section*{References}

%References are to be listed in the order cited in the text in Arabic
%numerals within square brackets. They can be
%referred to indirectly, e.g.~``$\ldots$
%in the statement \cite{beeson}.'' or used directly,
%e.g.~``$\ldots$ see [2] for examples.'' List references
%using the style shown in the following examples. For journal names,
%use the standard abbreviations.  Typeset references in 9 pt roman.

\end{document}